%% file: AHH-EPS.tex
\documentclass[a4paper,11pt]{article}
\usepackage{pos}
\usepackage{subcaption}
\usepackage{multirow}
\usepackage{cancel}

\input{paperdef}

\graphicspath{{figs/}}

\title{Large triple Higgs couplings in the 2HDM at $e^+e^-$ colliders}
\author*[a,b]{F.~ Arco}
\author[b]{S.~ Heinemeyer}
\author[a,b]{M.J.~Herrero}

\affiliation[a]{Departamento de F\'isica Te\'orica, 
Universidad Aut\'onoma de Madrid, 
Cantoblanco, 28049, Madrid, Spain}

\affiliation[b]{Instituto de F\'isica Te\'orica (UAM/CSIC), 
Universidad Aut\'onoma de Madrid, 
Cantoblanco, 28049, Madrid, Spain}

\emailAdd{francisco.arco@uam.es}
\emailAdd{Sven.Heinemeyer@cern.ch}
\emailAdd{maria.herrero@uam.es}

\abstract{Within the framework of the $\cp$ conserving Two
Higgs Doublet Models (2HDM) type I and II we investigate
the di-Higgs production at future $e^+e^-$ colliders in order to
find effects coming from triple Higgs couplings.
We define and explore some benchmark planes
that show large values of these couplings, still in agreement
with all the relevant theoretical and experimental constraints.
Within those planes two production channels are considered:
$e^+e^-\to h_ih_jZ$ and $e^+e^-\to h_ih_j\nu\bar{\nu}$,
with $h_ih_j=hh,\ HH\ \mathrm{and}\ AA$.
We discuss on the sensitivity to $\kala:=\lahhh/\lahhh^\mathrm{SM}$ and $\lahhH$ in the $hh\nu\bar{\nu}$ production and
to $\lahHH$ ($\lahAA$) in the $HH\nu\bar{\nu}$ ($AA\nu\bar{\nu}$) production at CLIC\,3\,TeV
via the cross section distribution on the invariant mass of the final-state Higgs-pair.
}

\FullConference{%
  *** The European Physical Society Conference on High Energy Physics (EPS-HEP2021), ***\\
  *** 26-30 July 2021 ***\\
  *** Online conference, jointly organized by Universität Hamburg and the research center DESY ***
}

\begin{document}
\begin{flushright}
\mbox{}
IFT--UAM/CSIC-21-118
\end{flushright}
\maketitle

\section{Introduction}
%In 2012, the ATLAS and CMS detectors finally discovered the last 
%building block of the Standard Model (SM): the Higgs boson.
%\cite{Aad:2012tfa,Chatrchyan:2012xdj,Khachatryan:2016vau}.
%The following measurements on this Higgs boson
%does not lead to any significant deviation from the SM. However, the precision
%on them, of around the 20\%, still leaves room for beyond SM (BSM) physics.

One of the most popular extensions of the SM is the two Higgs doublet
model (2HDM) \cite{Branco:2011iw}.
%In this contribution we only consider the 2HDM type I and type II.
This model leads to triple-Higgs boson interactions among 
the new scalars, namely $h$, $H$, $A$ and $H^\pm$,
which are the key focus of this contribution.
Future colliders, specially $e^+e^-$ colliders such as ILC \cite{Bambade:2019fyw}
and CLIC \cite{Charles:2018vfv}, will play a key role in order to discover (or exclude) beyond SM physics.
%, will improve 
%significantly the precision of the measurements of the Higgs couplings (see for instance
%\cite{deBlas:2019rxi, DiMicco:2019ngk}).
In this contribution, based on \citeres{Arco:2020ucn,Arco:2021bvf},
 we explore the effects of the triple Higgs couplings on the
production of two neutral Higgs bosons in the context of a 2HDM, type I and II,
at future $e^+e^-$ colliders.
The two principal production channels are $e^+e^-\to h_ih_jZ$ and $e^+e^-\to h_ih_j\nu\bar{\nu}$,
with $h_ih_j=hh,\ HH\ \mathrm{and}\ AA$.
We calculate the production rate of both channels in some benchmark planes and points
%, inspired by \cite{Arco:2020ucn}, 
and analyze the sensitivity to triple Higgs couplings 
via the cross section distribution on the invariant mass of the final-state Higgs-pair.
%We will focus on CLIC 3 TeV.

%For a full list of references see \cite{Arco:2020ucn,Arco:2021bvf}.

\section{The Two Higgs Doublet Model (2HDM)}
The 2HDM \cite{Branco:2011iw} contains two Higgs doublets, $\Phi_1$ and $\Phi_2$, 
contrary to the SM, where only one Higgs doublet is required.
In this contribution, we work in the context of a $CP$ conserving 2HDM with an imposed $Z_2$ symmetry,
in order to avoid flavor changing neutral currents, that is only softly broken by the parameter $\msq$.
This $Z_2$ symmetry leads to four different types of Yukawa sectors, where in this contribution we only consider 
2HDM type I and type II.
%\begin{eqnarray}
%V &=& m_{11}^2 (\Phi_1^\dagger\Phi_1) + m_{22}^2 (\Phi_2^\dagger\Phi_2) - \msq (\Phi_1^\dagger
%\Phi_2 + \Phi_2^\dagger\Phi_1) + \frac{\la_1}{2} (\Phi_1^\dagger \Phi_1)^2 +
%\frac{\la_2}{2} (\Phi_2^\dagger \Phi_2)^2 \nonumber \\
%&& + \la_3
%(\Phi_1^\dagger \Phi_1) (\Phi_2^\dagger \Phi_2) + \la_4
%(\Phi_1^\dagger \Phi_2) (\Phi_2^\dagger \Phi_1) + \frac{\la_5}{2}
%[(\Phi_1^\dagger \Phi_2)^2 +(\Phi_2^\dagger \Phi_1)^2]  \;,
%\label{eq:scalarpot}
%\end{eqnarray}
%where the $Z_2$ symmetry is only softly broken by the parameter $\msq$.
After the EW symmetry breaking, due to the new degrees 
of freedom, five physical states are realized: two $\cp$-even scalar fields,
$h$ and $H$, one $\cp$-odd one, $A$, and one charged pair, $H^\pm$.

We study the 2HDM in the physical basis, where the free parameters
%in \refeq{eq:scalarpot} can be expressed in terms of the following set of 
%parameters:
of the model are:
\begin{equation}
\Mh,~ m,~ \tb,~ \cos(\be-\al)\equiv
\CBA~ \mathrm{and}~ \msq,
\label{eq:inputs}
\end{equation}
where we consider a fully degenerated scenario with
$\MH=\MA=\MHp\equiv m$. The parameter $\tb$ is defined as the ratio of the vevs of the two Higgs doublets,
$\tb:=v_2/v_1$, satisfying the relation $v_1^2+v_2^2=v^2$, where $v$ is the SM vev.
The angles $\alpha$ and $\beta$ diagonalize the $\cp$-even and the $\cp$-odd
Higgs boson sectors respectively.
%The relations between the physical masses of the Higgs bosons and the
%couplings in \refeq{eq:scalarpot} can be found in \citere{Arco:2020ucn}.
In our analysis we identify the lightest $\cp$-even Higgs boson
$h$ with the one observed at $m_h \sim 125 \gev$ and the remaining Higgs bosons are 
assumed to be heavier. The limit $\CBA\to0$ is of great interest in the 2HDM because in that situation
all SM couplings for $h$ are recovered, what is known as the \textit{alignment limit}.
%However, in this limit not all BSM interactions vanish, couplings such as $hhH$ and $ZHA$
%are non-zero in the alignment limit. 
Regarding the constraints on the model, we consider theoretical constraints,
such as the tree-level perturbative unitarity and the stability of the potential, as well as 
experimental constraints, namely electroweak precision data, % (specially from the $T$ parameter),
the BSM boson searches by colliders, the properties of the 125 GeV Higgs boson %measured at the LHC and Tevatron 
and flavor observables. % (in particular the decays $B\to X_s\gamma$ and $B_s\to\mu\mu$). 
%It is worthy to mention that, contrary to our previous analysis in \citere{Arco:2020ucn},
%in this analysis the ``STXS observables'' given by the LHC are now included and this results in
%substantially stronger limits on $\CBA$, particularly in the 2HDM type~II.
A detailed discussion on these constraints including all references can be found in \cite{Arco:2020ucn,Arco:2021bvf}.

The key role in this contribution is played by the triple Higgs couplings $\la_{h_i h_j h_k}$.
We define
these couplings such that the Feynman rule of the interaction $h_i h_j h_k$ is given by
$
	- i\, v\, n!\; \la_{h_i h_j h_k}
$,
where $n$ is the number of identical particles in the vertex. 
We furthermore define $\kala := \lahhh/\lahhh^\mathrm{SM}$. 
The allowed range of triple Higgs couplings involving at least one SM-like
Higgs boson by all the relevant theoretical and experimental constraints can be found
in \cite{Arco:2020ucn}.

\section{\boldmath{$hh$} production in 2HDM type I}
The cross section predictions presented in this contribution
are calculated at tree-level 
 with the help of the public codes
{\tt{MadGraph5}}~\cite{Alwall:2014hca},
{\tt{FeynRules}}~\cite{Alloul:2013bka},
 {\tt 2HDMC} \cite{Eriksson:2009ws}
 and {\tt ROOT}~\cite{Brun:1997pa}. 
Electrons and positrons were assumed to be massless
and all diagrams are included in the computation. Here we
only consider production at $\sqrt{s}=3\tev$ (the highest energy potentially reachable at CLIC)
with an expected luminosity of $\cL_{\rm int}=5\mathrm{ab}^{-1}$ \cite{Charles:2018vfv}.
For other energies see \cite{Arco:2021bvf}.

%%%%%%%%%%%%%%%%%%%%%%%%% F I G U R E %%%%%%%%%%%%%%%%%%%%%%%%%%%%%%%%%%%%%%%%%
\begin{figure}[t!]
%\vspace{-2em}
\begin{center}
\begin{subfigure}{0.7\textwidth}
\includegraphics[width=0.48\textwidth]{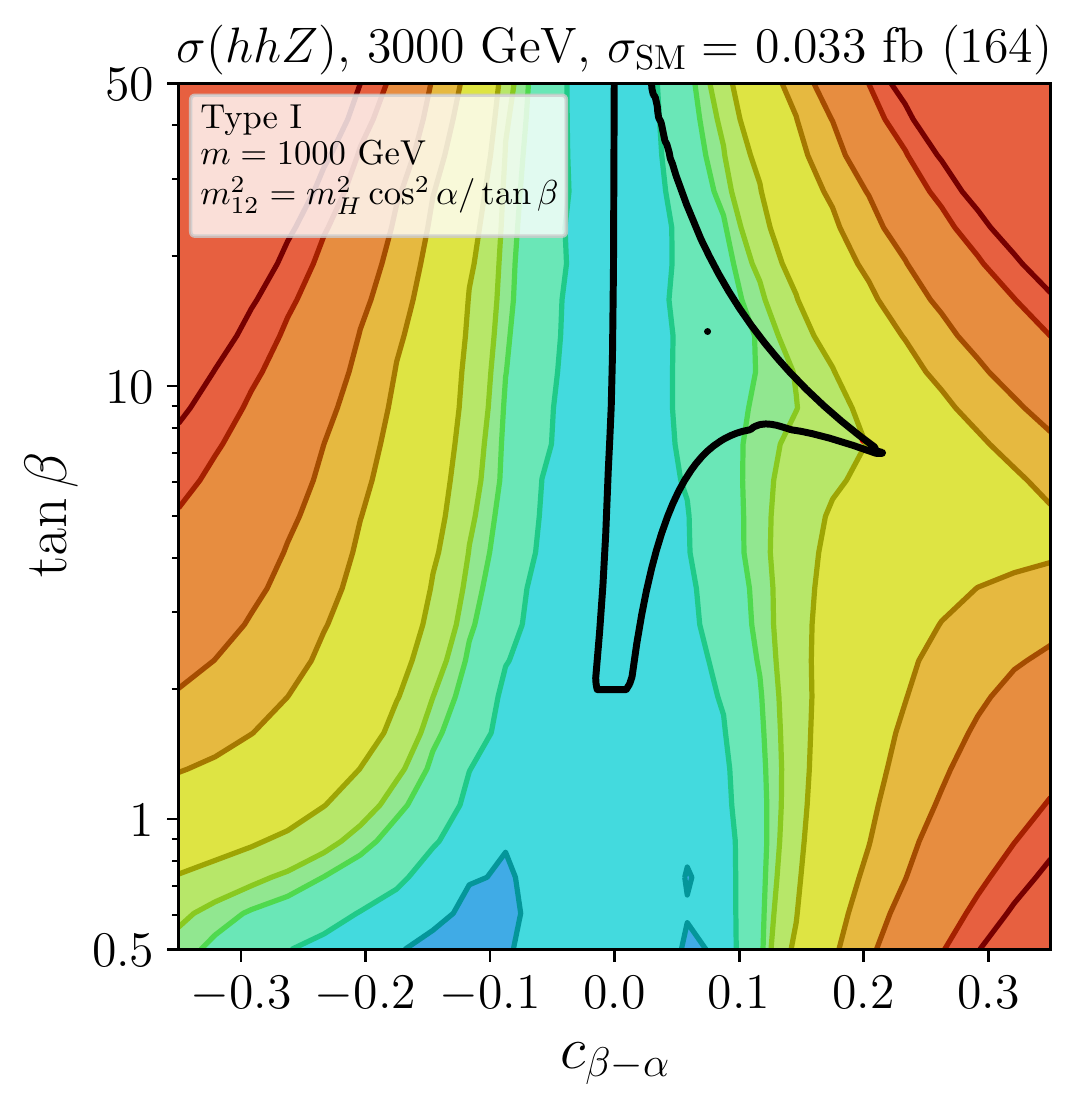}
\includegraphics[width=0.48\textwidth]{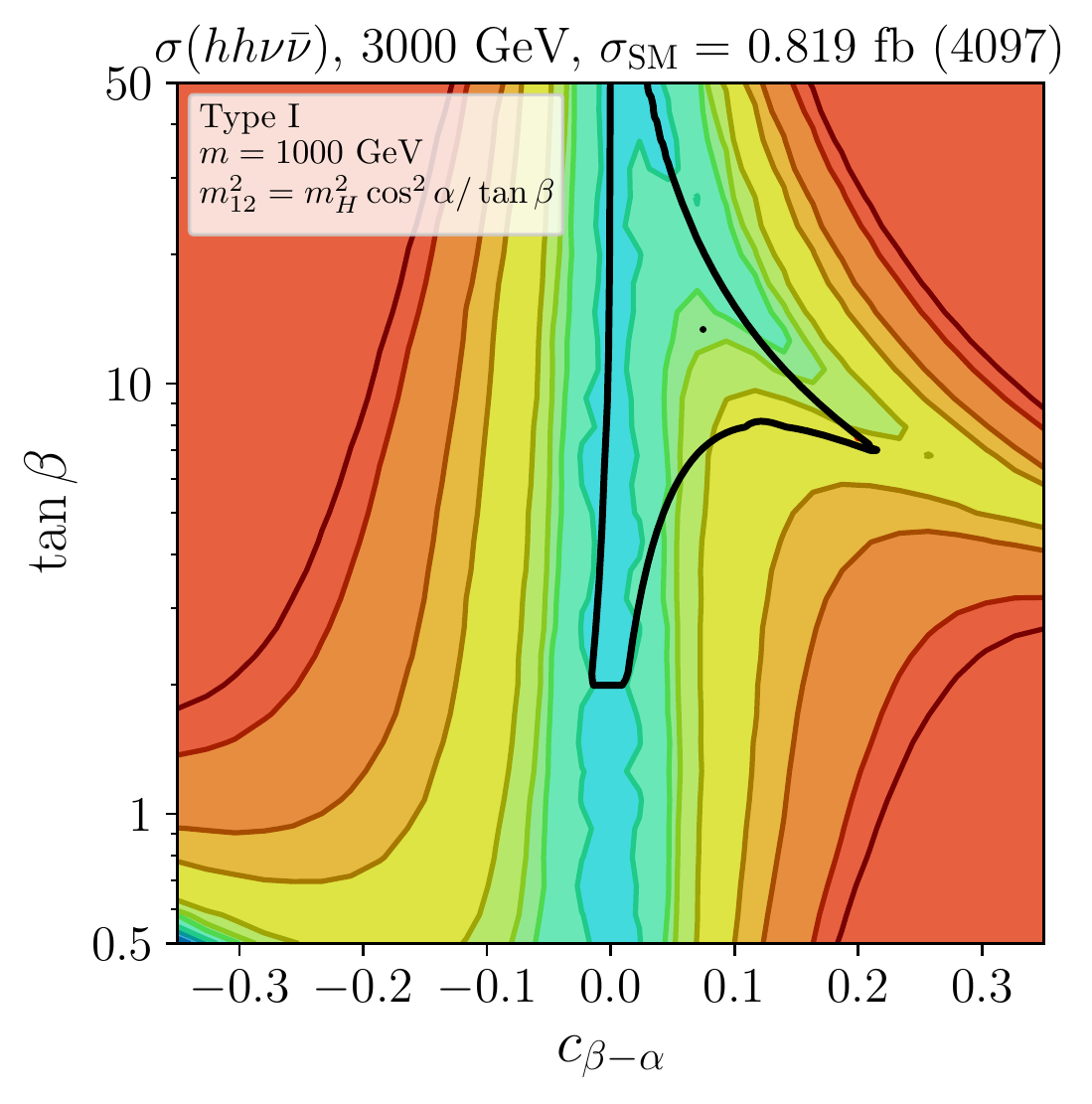}

\includegraphics[width=0.48\textwidth]{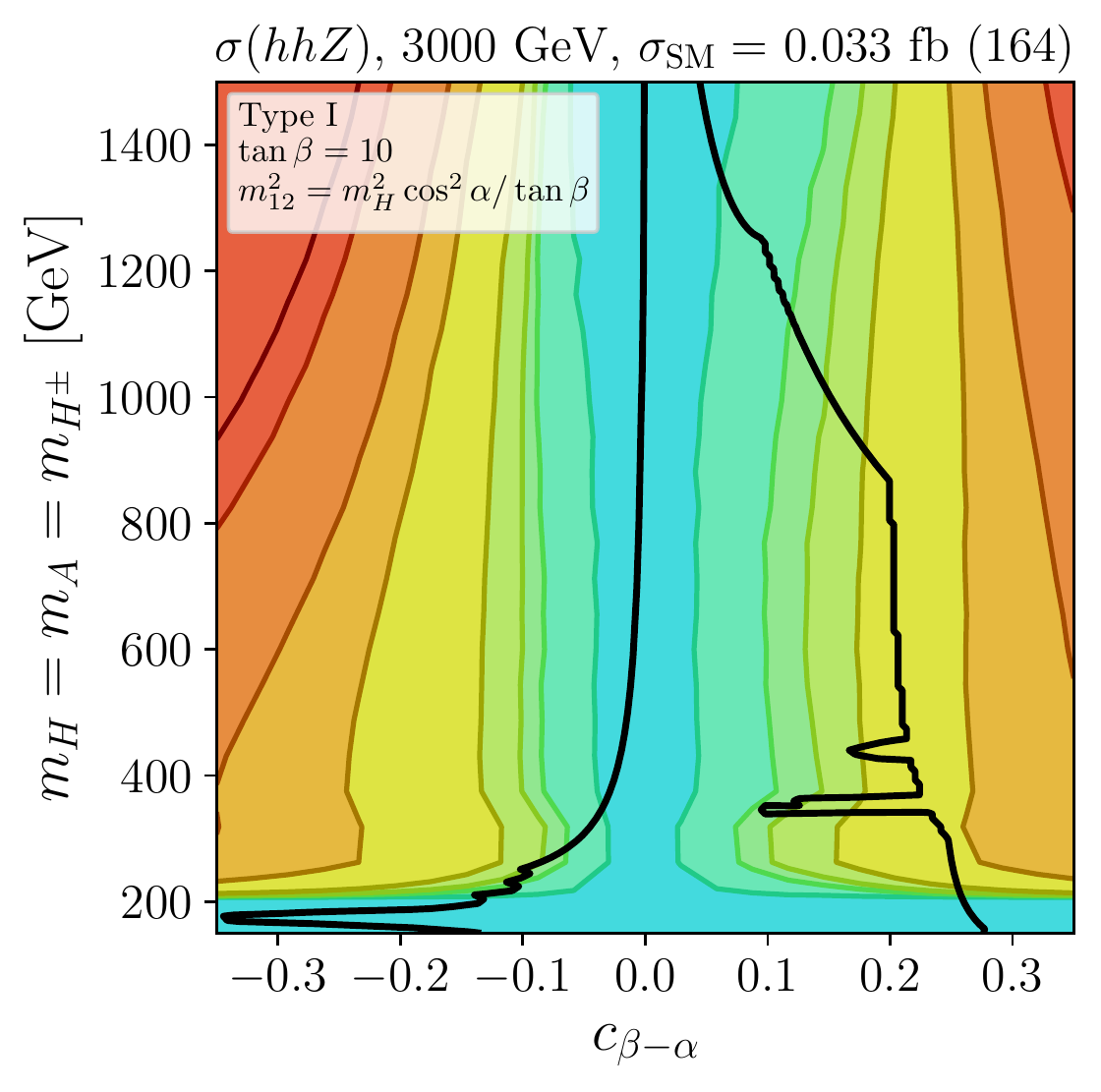}
\includegraphics[width=0.48\textwidth]{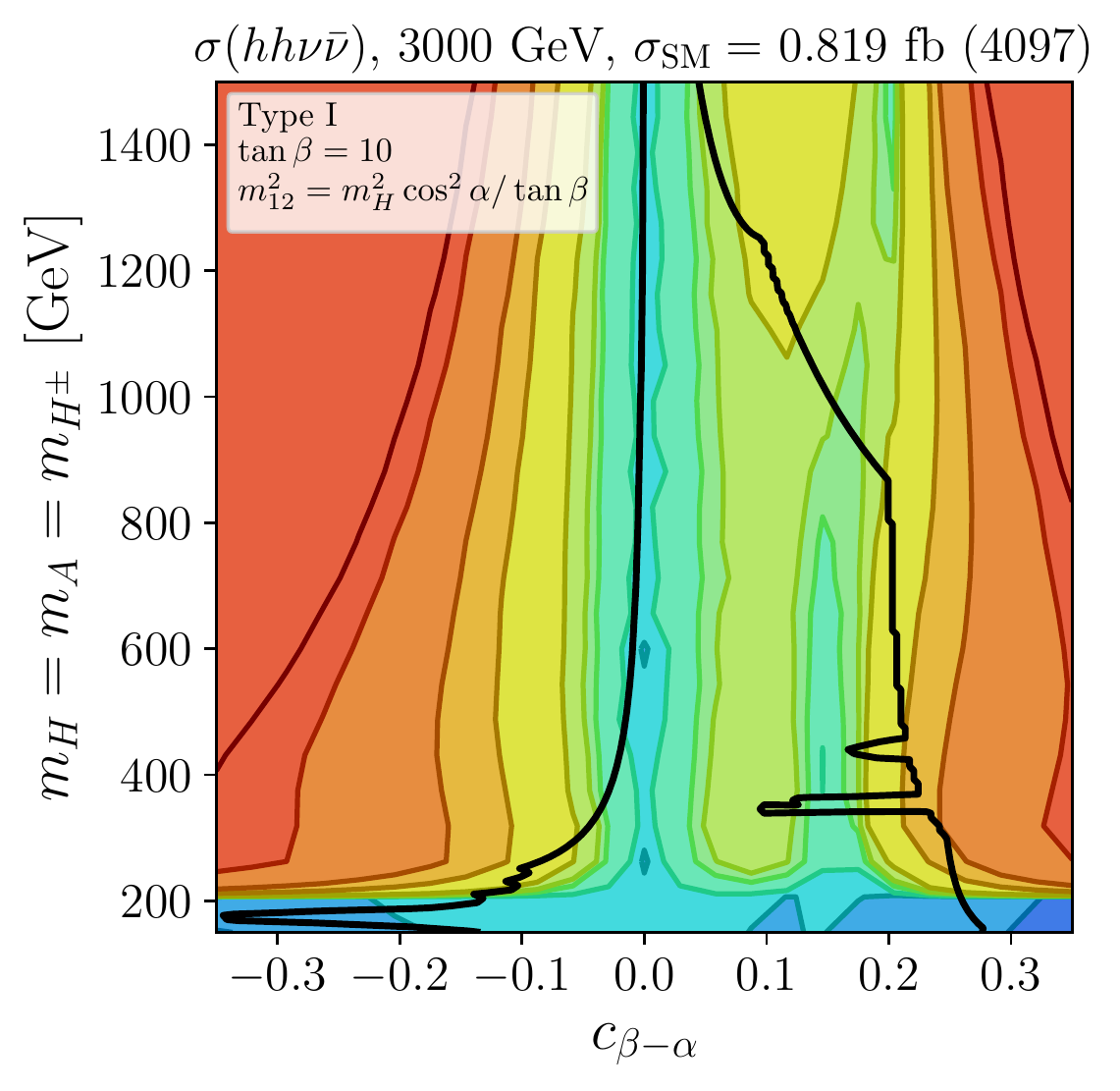}
\end{subfigure}	
\begin{subfigure}{0.2\textwidth}
\includegraphics[height=0.4\textheight]{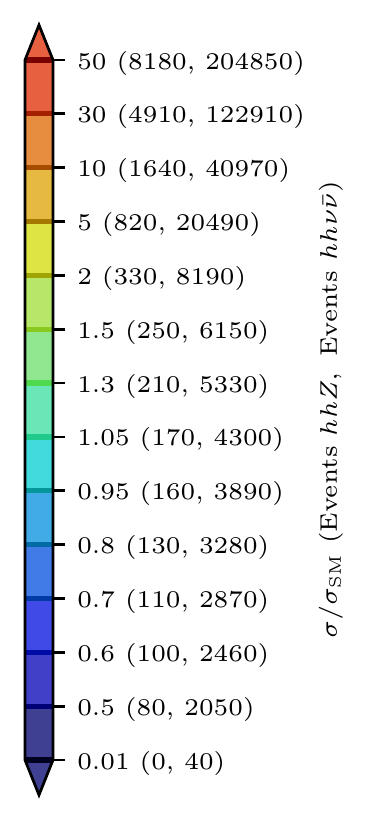}
\vspace{0.13cm}
\end{subfigure}
\end{center}
\vspace{-1em}
\caption{Cross sections for $e^+e^-\to hhZ$ (left) and
  $e^+e^-\to hh\nu\bar{\nu}$ (right) relative to the SM at $\sqrt{s}=3\tev$
  for the benchmark planes ($\CBA,\ \tb$) in the upper row and ($\CBA,\ m$) in the lower row.} 
\label{fig:xs_hh_3000-I}
\end{figure}
%%%%%%%%%%%%%%%%%%%%%%%%% F I G U R E %%%%%%%%%%%%%%%%%%%%%%%%%%%%%%%%%%%%%%%%%
%%%%%%%%%%%%%%%%%%%%%%%%% F I G U R E %%%%%%%%%%%%%%%%%%%%%%%%%%%%%%%%%%%%%%%%%
\begin{figure}[t!]
\begin{center}
\includegraphics[width=0.44\textwidth]{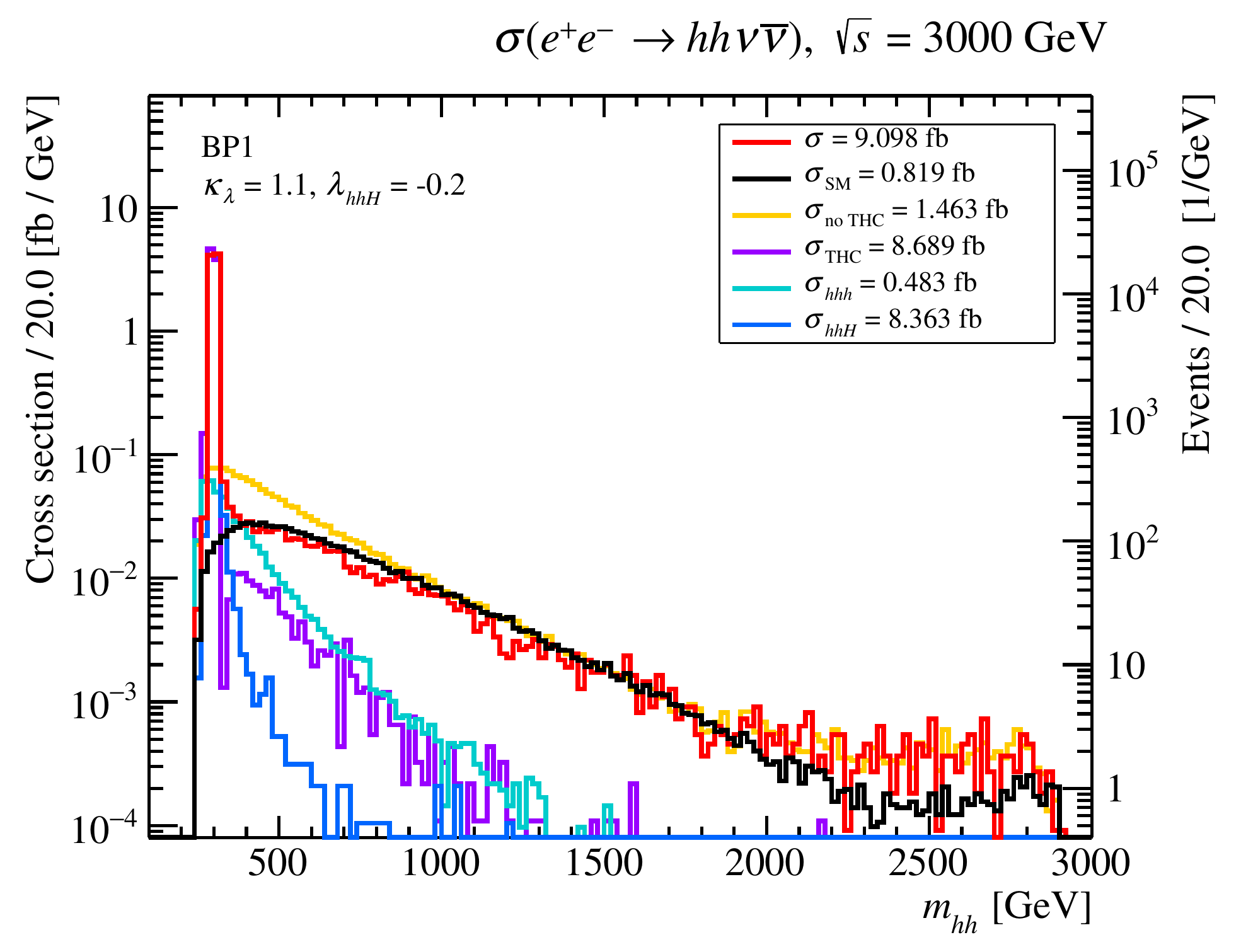}
\includegraphics[width=0.44\textwidth]{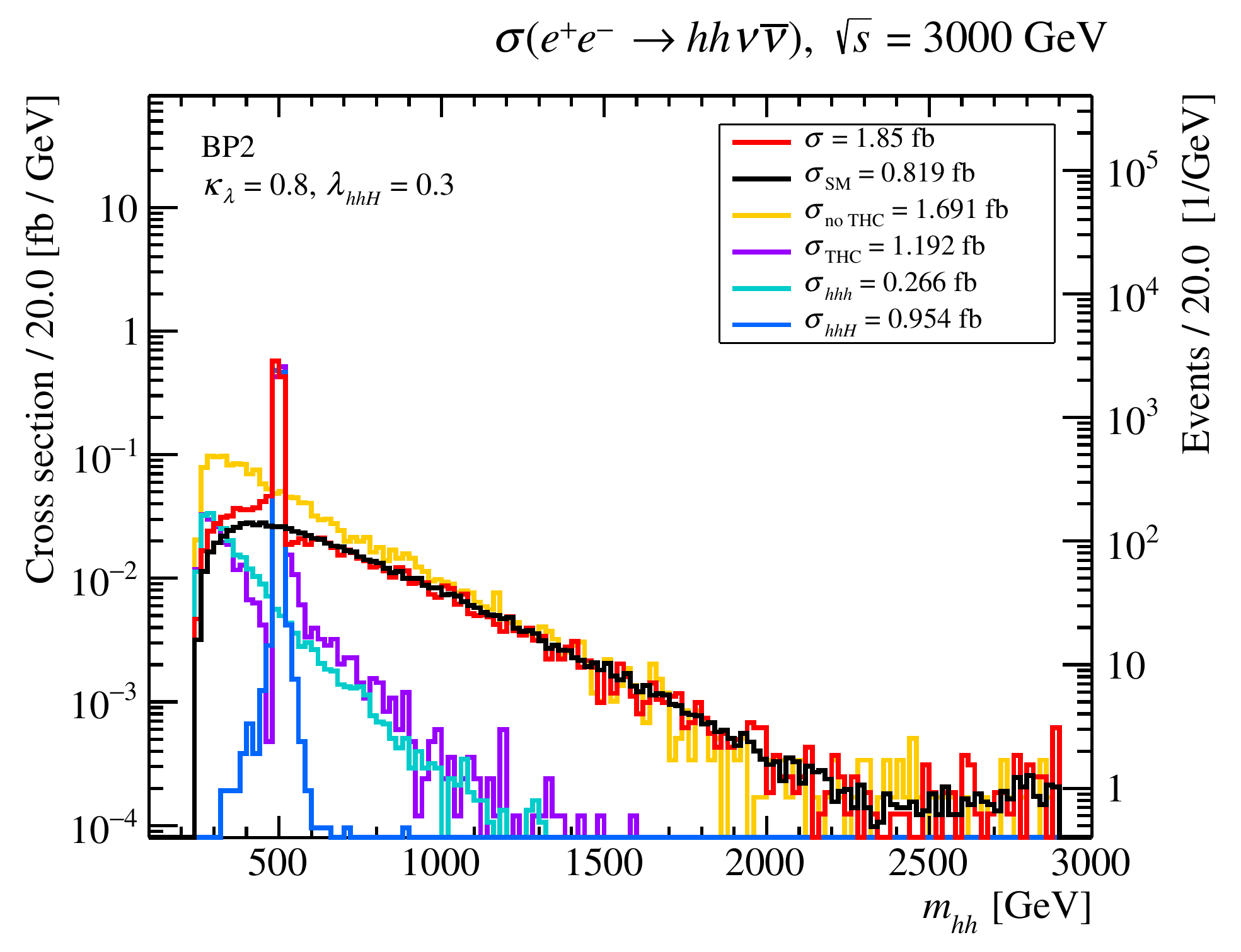}
\includegraphics[width=0.44\textwidth]{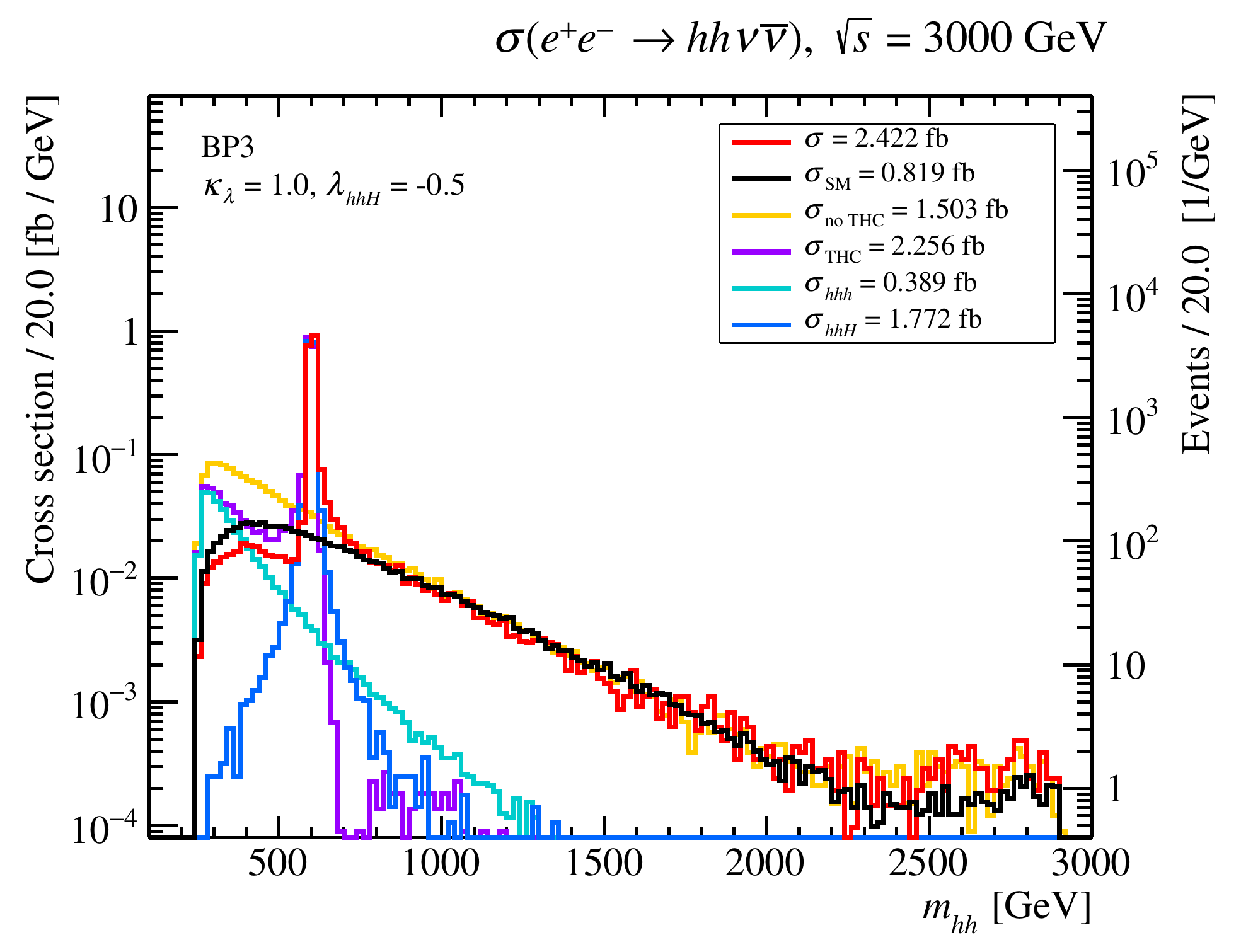}
\includegraphics[width=0.44\textwidth]{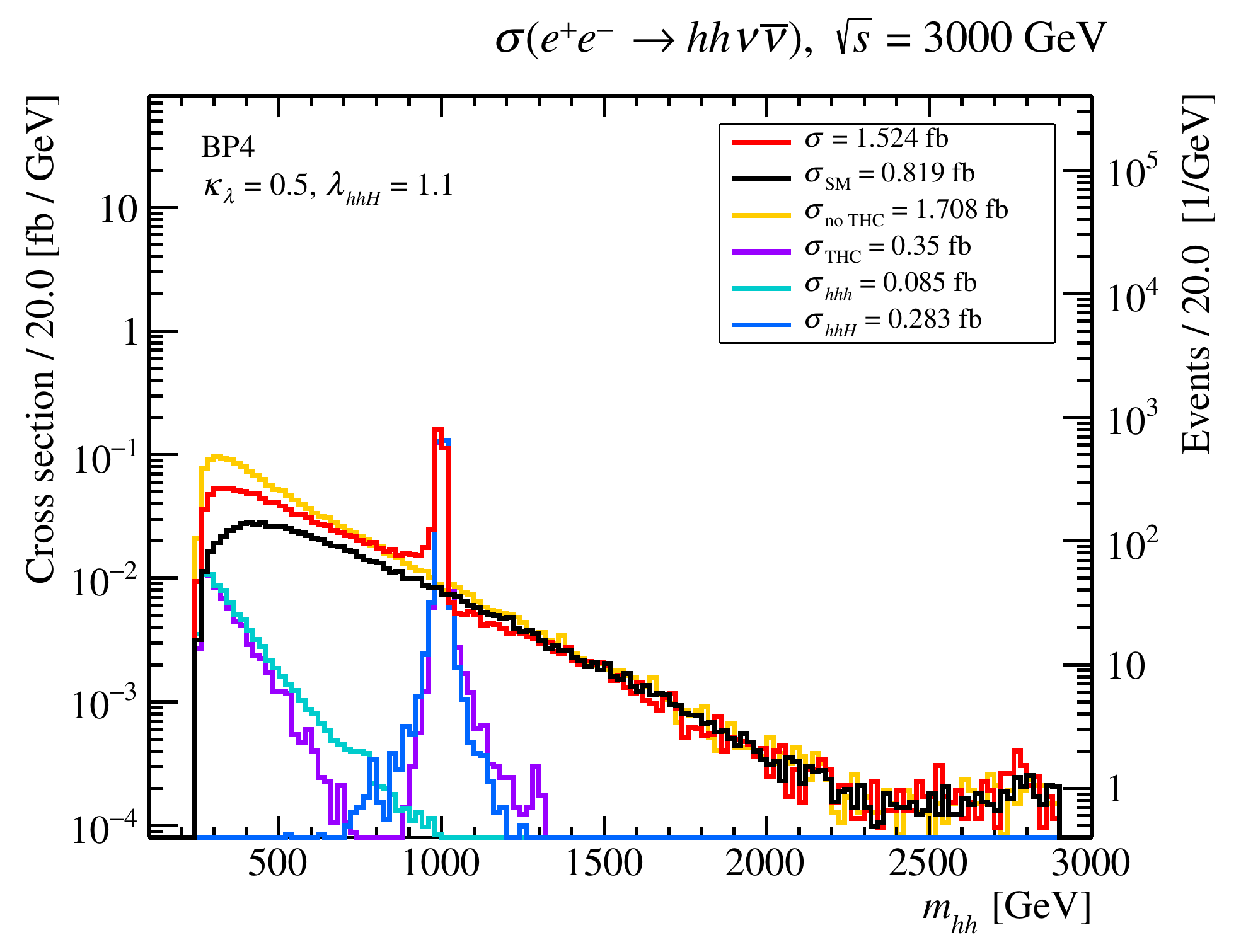}
\end{center}
\vspace{-1em}
\caption{Distribution on the invariant mass of the final-state $hh$ pair in the 
process $e^+e^-\to hh\nu\bar{\nu}$ at $\sqrt{s}=3\tev$ for BP1, BP2, BP3 and BP4 defined in \refta{tab:BP}.}
\label{fig:mhh3000}
%\label{fig:...}
\end{figure}
%%%%%%%%%%%%%%%%%%%%%%%%% F I G U R E %%%%%%%%%%%%%%%%%%%%%%%%%%%%%%%%%%%%%%%%%

In \reffi{fig:xs_hh_3000-I} we show the cross section predictions w.r.t.\ the SM prediction
for $e^+e^-\to hhZ$ and $e^+e^-\to hh\nu\bar{\nu}$ 
in the following benchmark planes in the 2HDM type I: plane ($\CBA,\ \tb$), with
  $m = 1 \tev$ and
  $\msq=\MH\cos^2\al/\tb$, and
  plane ($\CBA,\ m$), with
  $\tb = 10$ and
  $\msq=\MH\cos^2\al/\tb$.  
%Both planes present sizable triple Higgs couplings, in particular
%for sizable values for the ones present in the $hh$ production
%($\kala$ and $\lahhH$).
At this energy the neutrino
channel has a much larger cross section than the $Z$ channel.
In the ($\CBA,\ \tb$) plane (upper plots), cross sections around $2 \sigma_\mathrm{SM}$ can be found 
inside the allowed region (enclosed by the black solid line). On the other hand,
in the ($\CBA,\ m$) plane (lower plots) a cross section $\sim3\sigma_\mathrm{SM}$
can be realized for a wide range of heavy Higgs bosons masses within the allowed region.
Furthermore, $hh\nu\bar{\nu}$ production rates up to $10\sigma_\mathrm{SM}$ 
can be found inside the allowed region for very low $m$ and large $\CBA$.
%\subsection{Sensitivity to triple Higgs couplings}
In order to find sensitivity to triple Higgs couplings we propose the
study of the cross section distributions on the invariant mass of the final-state Higgs-pair.
In \reffi{fig:mhh3000}, we present the $\mathrm{d}\sigma/\mathrm{d}m_{hh}$ distribution for 
$e^+e^-\to hh\nu\bar{\nu}$ of four specific
benchmark points (BPs), whose input parameters are given in \refta{tab:BP}.
This set of BPs satisfies all the relevant constraints, % listed in the previous section,
covers a wide range on the heavy Higgs bosons masses and
leads to sizable triple Higgs couplings.
%In addition, the cross section distributions are presented by separated 
%contributions concerning different set of diagrams: the complete 2HDM
%distribution (red lines), the contribution from diagrams without triple Higgs couplings
%(yellow lines), the contribution from diagrams with $\kala$ (light blue lines), 
%the contribution from diagrams with $\lahhH$ (dark blue lines), 
%the contribution from the last two (purple lines). It is also displayed the
%complete SM distribution for comparison (black lines).
%%%%%%%%%%%%%%%%%%%%%%%%% T A B L E %%%%%%%%%%%%%%%%%%%%%%%%%%%%%%%%%%%%
\begin{table}[t!]
\begin{center}
{\small
\begin{tabular}{c|c|c|c|c|c|c|c|c|c|c}
Point & $m$  & $\tan\beta$ & $c_{\beta-\alpha}$ & $\msq$  & $\kala$ & $\lahhH$ & $\lahHH$ & $\laHHH$ & $\Gamma_H$  & $\Gamma_A$ \tabularnewline
\hline 
BP1  & 300  & 10 & 0.25 & $\MH\cos^2\al/\tb$ & 1.1 
& -0.2 & 2 & 0.3 & 0.84 & 0.18 \tabularnewline
\hline 
BP2 & 500  & 7.5 & 0.1 & 32000 & 0.8 & 0.3 & 2 & 0.6 & 0.88 & 0.71 \tabularnewline
\hline 
BP3 & 600  & 10 & 0.2 & $\MH\cos^2\al/\tb$  & 1.0 & -0.5 & 6 & 0.6 & 5.1 & 2.6 \tabularnewline
\hline 
BP4 & 1000  & 8.5 & 0.08 & $\MH\cos^2\al/\tb$  & 0.5 & 1.1 & 6 & -0.2 & 5.8 & 2.6
\end{tabular}}
\end{center}
\vspace{-1.5em}
\caption{Benchmark points in the 2HDM type I (masses and widths are given in GeV). \hspace{3cm} 
} 
\label{tab:BP}
\end{table}
%%%%%%%%%%%%%%%%%%%%%%%%% T A B L E %%%%%%%%%%%%%%%%%%%%%%%%%%%%%%%%%%%%
It can be seen in \reffi{fig:mhh3000} that the effect from $\kala$ (light blue lines) for all BPs appears
close to the threshold region at $m_{hh}=250\gev$, similar to
what happens in the SM. Therefore, we expect that the
sensitivity to $\kala$ to be comparable to that obtained for
future $e^+e^-$ colliders in the context of the SM \cite{Roloff:2019crr}. 
%One special situation can occur when $m_H$ is not much larger than $2 m_h$ and the
%contribution from the diagrams mediated by an intermediate $H$ boson can
%give a sizable contribution in the threshold region, just what happens in BP1.
Furthermore, from the comparison of the distributions of the
total cross section (red lines) and the contribution from diagrams without 
triple Higgs couplings (yellow lines) that close to threshold region
one can deduce that diagrams with $\kala$ have a destructive interference
with the diagrams without triple Higgs couplings
 in the $hh\nu\bar{\nu}$ production, just as in the SM.
The sensitivity to $\lahhH$ enters through $H$ mediated diagrams (dark blue lines) that
can produce a resonant peak around $m_{hh}=m_H$,
as it can be seen in all BPs in \reffi{fig:mhh3000}. 
%A sensitivity to the sign of $\lahhH$ can also be extracted,
%since the resonant peaks presents an asymmetry due to interference of the $H$
%mediated diagrams with the rest of the contributions and the change of sign the suffers
%this kind of diagrams.
We propose a theoretical estimator $R$ to study the sensitivity to $\lahhH$ 
via the $h\to b\bar b$ decays defined as:
\begin{equation}
R=\frac{{\bar N}^R-{\bar N}^C}{\sqrt{{\bar N}^C}}~,\ \ \mathrm{with}\ \ \bar N =N \times {\cal A} \times (\epsilon_b)^4,
\end{equation}
where $N$ is the number of final four $b$-jets events nearby the resonant peak,
the superscript $R,C$ refers to events coming from the $H$ mediated resonant diagrams
and from the diagrams with no triple Higgs couplings, respectively.
$\epsilon_b$ is the $b$-tagging efficiency of the detector, 
that we considered around 80\%, % (see, e.g., 
%\cite{Asner:2013psa,Abramowicz:2016zbo})
and 
$\cal A$ is the detection acceptance assuming the following cuts
(see \cite{Arco:2021bvf} for details):
\begin{equation}
\label{cuts}
p_T^b>20\enspace \mathrm{GeV} ;\,\,\,\,
\vert\eta^b\vert<2 ;\,\,\,\,
\Delta R_{bb}>0.4 ;\,\,\,\,
p_T^Z>20\enspace \mathrm{GeV} ;\,\,\,\,º
\cancel{E}_T>20\enspace \mathrm{GeV}.
\end{equation} 
We find larger values of $R$ for the $hh\nu\bar\nu$ production channel than for
$hhZ$ at all studied center of mass energies, except for BP1 at $\sqrt{s}=500\gev$ \cite{Arco:2021bvf}.
Moreover, the larger values of $R$ are found for $hh\nu\bar\nu$ at 3 TeV.
Therefore, CLIC\,3\,TeV %(the latest and most energetic projected stage)
 is the best suited collider to access to $\lahhH$ according to our estimator $R$.
The precise values for $\bar N^{R/C}$, $\cal A$ and $R$ for all BPs for both 
$hhZ$ and $hh\nu\bar\nu$ can be found in \cite{Arco:2021bvf}.

\section{\boldmath{$HH\nu\bar\nu\sim AA\nu\bar\nu$} production in 2HDM type II}
%In this section we study the cross section production of $HH\nu\bar\nu$ and $AA\nu\bar\nu$
%in the 2HDM type II.
%Since the collider constraints force the 2HDM type II to be very close to the alignment limit, 
%the $HH$ production is the only that can produce interesting situations, because
%if $\CBA=0$ then the $hh$ production in the 2HDM is the same as in the SM and the $hH$
%production is zero.
%In the degenerated scenario where $m_H=m_A$, we have that $HH$ and $AA$ production
%are numerically equivalent, so in this section we will only refer to the first one.

\begin{figure}[t!]
	\begin{center}
\includegraphics[width=0.35\textwidth]{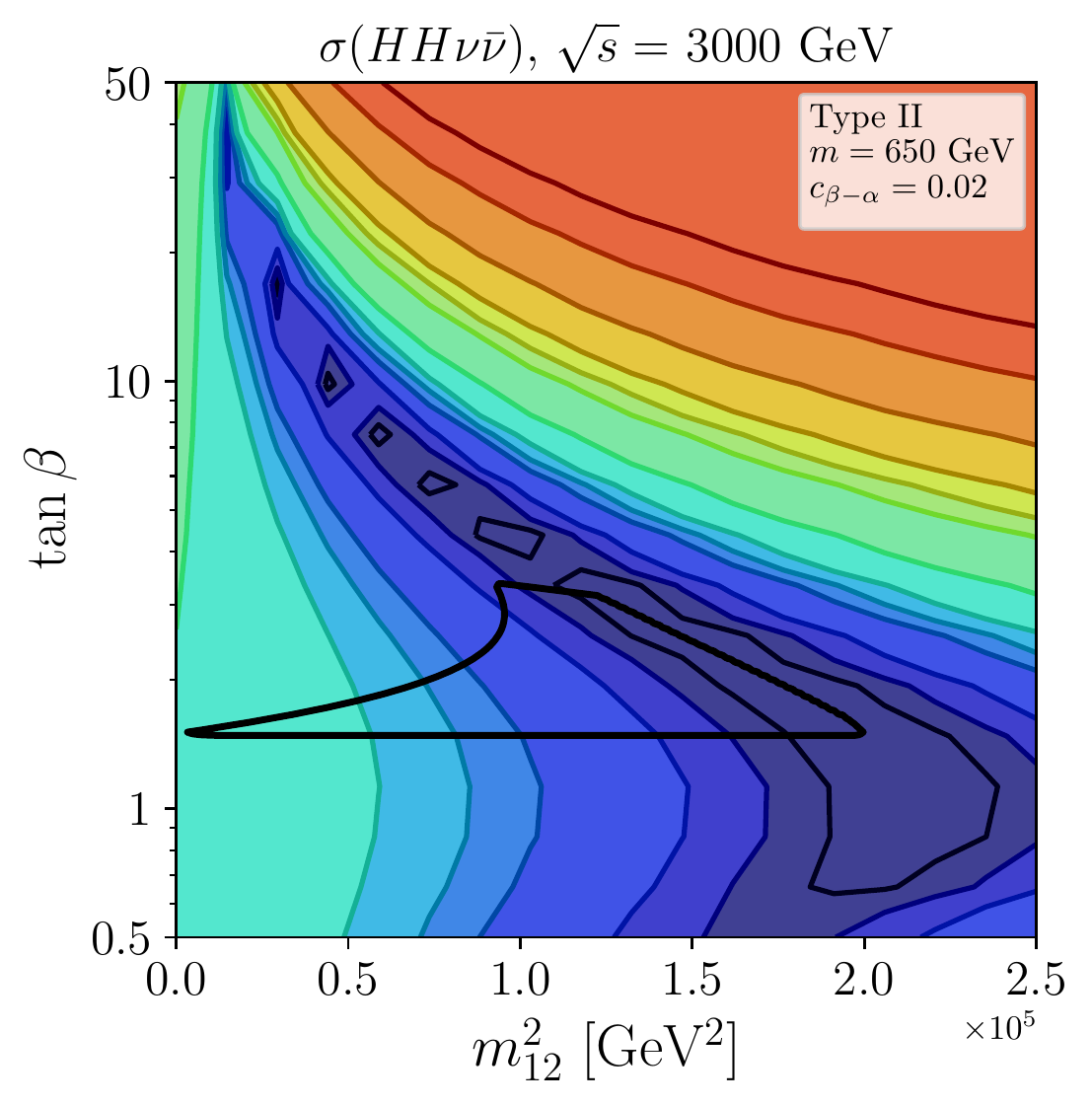}
\includegraphics[width=0.12\textwidth]{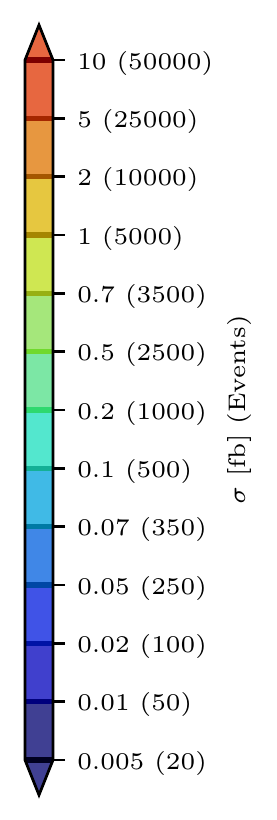}
\includegraphics[width=0.44\textwidth]{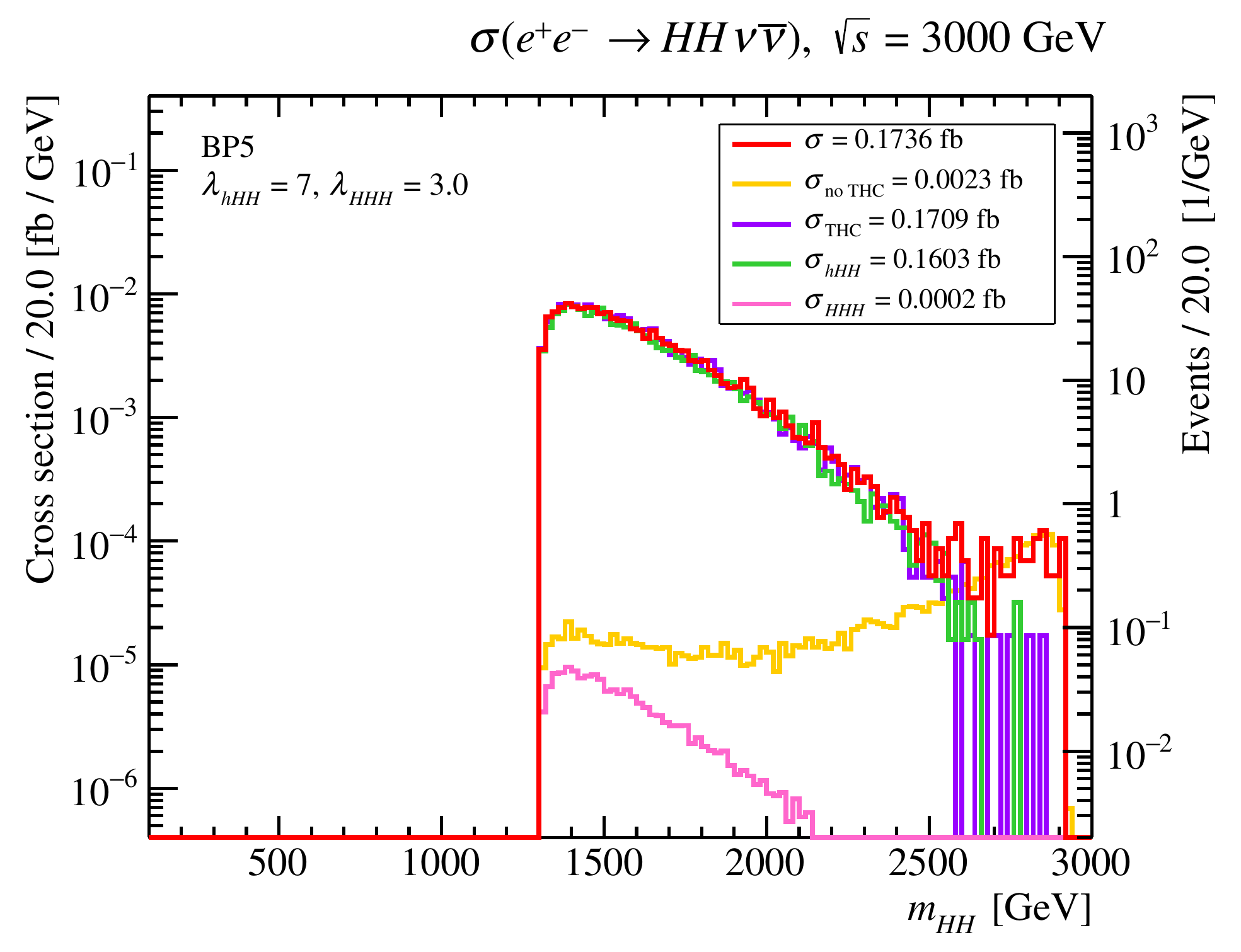}
	\end{center}
\vspace{-1em}
\caption{Cross section for $e^+e^-\to HH\nu\bar{\nu}$ (left)  and
distribution on the invariant mass of the final-state $HH$ pair for BP5 (right) at $\sqrt{s}=3\tev$.
}
\label{fig:typeII}
%\label{fig:...}
\end{figure}

In the left panel of \reffi{fig:typeII} we show the production cross section of $HH\nu\bar\nu$
at 3 TeV in the plane ($\msq,\ \tb$) with
$m =650\gev$ and $\CBA=0.02$, in 2HDM type II. In the degenerated scenario
where $m_H=m_A$ one finds $\sigma(HH\nu\bar\nu)\simeq\sigma(AA\nu\bar\nu)$,
so we only show the former.
It can be seen that the cross section reaches its maximum value, close to $0.2\fb$, 
within the allowed region (delimited by the solid black line) for $\msq\lesssim 5\times10^5\gev^2$.
This maximum coincides with the largest values for $\lahhH$ in that plane,
where couplings around 7 are realized. 
On the other hand, in the left panel of \reffi{fig:typeII} we show
the differential cross section distribution of $e^+e^-\to HH\nu\bar \nu$ as a function of
$m_{HH}$ for the point $\tb=1.5$ and $\msq=10000\gev^2$ within the previously discussed
plane (which is denoted as BP5). %The color code is as in \reffi{fig:mhh3000}, but here
%it is shown the contribution from diagrams containing $\lahHH$ (green lines) and the 
%contribution from diagrams containing $\laHHH$ (pink lines).
One can see that the effect coming from the diagrams depending on $\lahHH$ (green lines)
dominates over the rest of the contributions and is nearly fully responsible for the total
cross section. 
This indicates that the total cross section can receive a relevant
contribution from the diagrams containing $\lahHH$, in particular from
the non-resonant contributions, which dominate at this large center-of-mass energy. 
It is worth noticing that
this situation can be reproduced in the 2HDM type I if a similar value of $\lahHH$
is realized. 

%%%%%%%%%%%%%%%%%%%%%%%%% F I G U R E %%%%%%%%%%%%%%%%%%%%%%%%%%%%%%%%%%%%%%%%%
%%%%%%%%%%%%%%%%%%%%%%%%% F I G U R E %%%%%%%%%%%%%%%%%%%%%%%%%%%%%%%%%%%%%%%%%

\section{Summary and conclusions}
%An important task at future colliders is the investigation of the
%Higgs-boson sector. Here the measurement of the triple Higgs
%coupling(s) plays a special role.
In this contribution, based on \citeres{Arco:2020ucn,Arco:2021bvf},
we show the di-Higgs production cross sections
%w.r.t.\ triple Higgs couplings
at future high energy $e^+e^-$ colliders in
the 2HDM type I and II.
Here we focus on a center-of mass energy of 3 TeV, 
the projected final energy state of CLIC.
We consider two different channels,
$e^+e^- \to h_i h_j Z$ and $e^+e^- \to h_i h_j \nu \bar\nu$ with 
$h_i h_j=hh,HH,AA$.
%We find that cross section of the di-Higgs production together with a $Z$ boson (a neutrino-antineutrino pair)
%decreases (increases) with the collider energy, with the exception of the $hH\nu\bar\nu$ channel.
We analyze the production rate of these processes  
within several benchmark planes that present sizable values for triple Higgs
couplings while being in agreement with all theoretical and experimental constraints.
By means of the cross section distributions on  $m_{h_ih_j}$,
we find sizable effects from 
$\kala$ and $\lahhH$ in the $hh\nu\bar\nu$ production and 
effects from $\lahHH$ and $\lahAA$ in the $HH\nu\bar\nu$ and $AA\nu\bar\nu$ production
respectively.%, both at CLIC 3 TeV.
We find a 
dependence on $\kala$ similar to the one present in the SM, where the
largest influence on the cross section happens on the region slightly above $m_{hh}=250\gev$.
On the other hand, the dependence on $\lahhH$ enters through resonant diagrams mediated
by the $H$ boson. We found sensitivity to this resonance (and in consequence to $\lahhH$) 
for a wide range of masses of the BSM heavy Higgs bosons. Via a theoretical estimator
$R$, that takes into account the detection of the final four $b$-jet state, 
we conclude that CLIC\,3\,TeV
will be the most sensitive future $e^+e^-$ collider to $\lahhH$.
Finally, we find that $HH\nu\bar\nu$ ($AA\nu\bar\nu$) can be completely
dominated by the value of $\lahHH$ ($\lahAA$), 
reaching production rates up to 0.2 fb. That would mean that an experimental
detection of these channels would provide an interesting opportunity to measure $\lahHH$ and
$\lahAA$ at CLIC\,3\,TeV.

\end{document}

%% file: paperdef.tex
\newcommand\al{\alpha}
\newcommand\be{\beta}
\newcommand\tb{\tan\beta}

\newcommand\CBA{c_{\beta - \alpha}}

\newcommand\ReDiag{\mathop{%
  \raise .5pt\hbox{[}%
  \widetilde{\mathrm{Re}}%
  \raise .5pt\hbox{]}}}
\newcommand\ReOffDiag{\mathop{%
  \raise .5pt\hbox{$\llbracket$}%
  \widetilde{\mathrm{Re}}%
  \raise .5pt\hbox{$\rrbracket$}}}

\newcommand\cL{{\cal L}}

\newcommand\Mh{m_h}
\newcommand\MH{m_H}
\newcommand\MA{m_A}
\newcommand\MHp{m_{H^\pm}}

\newcommand\msq{m_{12}^{2}}

\newcommand\refta[1]{Tab.~\ref{#1}}

\newcommand\citeres[1]{Refs.~\cite{#1}}

\newcommand{\CP}{{\cal CP}}
\newcommand{\cp}{{\CP}}

\newcommand{\tev}{\,\, \mathrm{TeV}}
\newcommand{\gev}{\,\, \mathrm{GeV}}

\newcommand\fb{\ensuremath{\,\mbox{fb}}}

\def\reffi#1{\mbox{Fig.~\ref{#1}}}

\def\la{\lambda}
\newcommand\kala{\ensuremath{\kappa_{\lambda}}}

\newcommand{\lahhh}{\ensuremath{\la_{hhh}}}
\newcommand{\lahhH}{\ensuremath{\la_{hhH}}}
\newcommand{\lahHH}{\ensuremath{\la_{hHH}}}
\newcommand{\lahAA}{\ensuremath{\la_{hAA}}}

\newcommand{\laHHH}{\ensuremath{\la_{HHH}}}

\definecolor{Orange}{named}{orange}
\definecolor{Purple}{named}{purple}
\definecolor{Lightblue}{cmyk}{0.9,0.1,0.1,0.3}
\definecolor{dgelborange}{cmyk}{0.,0.3,0.5, 0.}
\definecolor{Lila}{rgb}{0.5,0.,1}
\definecolor{Darkgreen}{rgb}{0.,.7,0.2}